\documentclass[a4paper,12pt]{article}
\usepackage{amsfonts}
\usepackage{amsmath}

\input{tcilatex}
\begin{document}

\title{The 't Hooft coupling and baryon mass splitting in the large-$N$
chiral model}
\author{Duojie JIA$^{\dagger }\thanks{%
Email:jiadj@nwnu.edu.cn}$, JiaShen ZHANG$^{\dagger }$ \\
$^{\dagger }$Institute of Theoretical Physics, College of Physics and\\
Electronic Engineering, Northwest Normal University,\ \\
Lanzhou 730070, China}
\maketitle

\begin{abstract}
{\normalsize We study the 't Hooft coupling }$g_{t}${\normalsize \ and the
mass splitting of the ground-state baryons in a chiral quark model inspired
by the large }$N_{c}\ ${\normalsize QCD. Depending on }$g_{t}${\normalsize \
the Hartree wavefunction of light quark in baryons is used to map the mass
hyperfine splittings for the octet and decuplet baryons. The 't Hooft
coupling }$g_{t}${\normalsize \ that reproduces the data of baryon masses is
determined to be around }$1.57${\normalsize , which is small in the sense
that }$g_{t}^{2}/(4\pi N_{c})\ll ${\normalsize \ }$1${\normalsize . }

{\normalsize PACS number(s):12.20.Ds, 11.15.Tk, }14.20.-c,11.27.+d

\end{abstract}

\section{Introduction}

The 't Hooft's idea \cite{tHooft} that Quantum Chromodynamics (QCD)
simplifies in the limit of large number of colors, $N_{c}$, had proven to be
a valuable tool for exploring strong interaction\cite{WittenB,ColemanN}.
With the $1/N_{c}$ expansion of QCD, the qualitative explanations can be
given for, for instance, the Okubo-Zweig-Iizuka rule, the Regge
phenomenology, the soliton picture of baryon\cite{WittenB}, the contracted $%
SU(2n_{f})$ spin-flavor symmetry\cite{DashenMano} of QCD, and the
spin-flavor structure of many baryon properties\cite%
{Luty,Carone,DashenJenkinsM,PirjolYan,Schat}. The existences of the smooth
limit of QCD at $N_{c}\rightarrow \infty $, with $g^{2}N_{c}$ fixed($g$ is
the QCD coupling constant), has been studied and confirmed via lattice
simulation of $SU(N_{c})$ gauge theory\cite{Teper,Lucini,Debbio} (see also 
\cite{LSUN} for a recent review).

It is hard, however, to say on theoretical grounds whether $1/N_{c}$ $=1/3$
can be considered small in real world where $N_{c}=3$ without actually
solving QCD for finite $N_{c}$. This is so because the answer depends
crucially on how large are the coefficients companying with the expanding
parameter $1/N_{c}$, the 't Hooft's coupling $g_{t}\equiv g^{2}N_{c}$, which
is assumed to be fixed to ensure the $SU(N_{c})$ QCD(the one-loop gluon
vacuum polarization, for instance) to have a smooth limit as $N_{c}$ tends
to infinity. As argued recently by Weinberg\cite{Weinberg10}, the 't Hooft's
coupling $g_{t}$ may not be small at moderate energies such as the scale of $%
m_{\rho }$ since the masses of the $\rho $ mesons $m_{\rho }\ $is of order
of $\Lambda _{QCD}$ in the chiral limit. Combining the $1/N_{c}$ expansion
of QCD with the early idea of the chiral theory \cite%
{Weinberg79,ManoharGeorgi}, Weinberg proposed a large-$N_{c}$ chiral theory%
\cite{Weinberg10} of quarks, gluons, and pions which is effectively
renormalizable to leading order of $1/N_{c}$. He suggests that such an
effective field theory, with a small value of the `t Hooft coupling $g_{t}$
at moderate energies, may explain why the naive quark model are so
successful. It remains unknown how small $g_{t}$ could be in the effective
theory.

The purpose of this Letter is to estimate the value of the 't Hooft coupling 
$g_{t}$ in the quark picture of baryons with pion included. We determine, in
a large-$N_{c}$ chiral quark model that is inspired by the Weinberg's large-$%
N_{c}$ effective theory, the 't Hooft coupling by computing the mass
splittings of the ground-state octet and decuplet baryons with the help of
Witten's Hartree picture and mapping them to the data of baryon masses. The
't Hooft coupling $g_{t}$ is found to be about $g_{t}$ $\simeq 1.57$. This
infer that {\normalsize when pion enters in the quark picture} $g_{t}$%
{\normalsize \ can be considered small in the sense that }$g_{t}^{2}/(4\pi
N_{c})\ll ${\normalsize \ }$1$.{\normalsize \ }

The model we utilized here is a potential version of the large-$N_{c}$
effective theory\cite{Weinberg10} where the gluon interaction is effectively
described by the Hartree potentials with the form of the coulomb plus string
potential. Conceptually, it resembles the (topological) chiral bag model 
\cite{ChBag} in the sense that the role of bag is played by the string-like
confining potential, with the chiral symmetry implemented via the quark-pion
interaction.

\section{Confined quarks in baryon at large $N_{c}$}

Inspired by the large-$N_{c}$ QCD and the chiral theories\cite{Weinberg10},
we use a large-$N_{c}$ chiral model of baryon\cite{Weinberg79,ManoharGeorgi}
in which each of quark $q_{i}(i=1,\cdots ,N_{c})$ in baryon moves in the
confining Hartree potential of all $N_{c}-1$ others, coupling with the pion
in an invariant way under chiral symmetry. The model is 
\begin{equation}
\begin{array}{c}
\mathcal{L}^{Baryon}=\sum_{i=1}^{N_{c}}\bar{q}_{i}[i\gamma ^{\mu }\partial
_{\mu }-(m_{q}+S)U_{5}-\gamma ^{0}V_{G}]q_{i} \\ 
+\mathcal{L}^{\chi },%
\end{array}
\label{chQM}
\end{equation}%
where $m_{q}$ is the effective mass of the light (up or down) quark $q_{i}$
and ($S,V_{G},U_{5}$) are the Hartree fields experienced by $q_{i}$, created
by all other quarks in baryon. The scalar and vector interactions, $S$ and $%
V_{G}$, stand for the effective QCD gluon interaction, which can be
approximated by the string potential and the one-gluon-exchange(OGE)
interaction, while 
\begin{equation*}
U_{5}\equiv \exp (2i\gamma ^{5}\boldsymbol{T}\cdot \boldsymbol{\pi }/f_{\pi
}),
\end{equation*}%
is the chiral interaction that $q_{i}$ is involved, with $f_{\pi }$ the pion
decay constant. The chiral dynamics $\mathcal{L}^{\chi }$ of the pion mesons 
$\boldsymbol{\pi }$ is chosen, for simplicity, be the Skyrme Lagrangian\cite%
{Skyrme},%
\begin{equation}
\mathcal{L}^{\chi }=\frac{f{}_{\pi }^{2}}{4}tr(\partial _{\mu }U\partial
^{\mu }U^{\dag })+\frac{1}{32e_{s}^{2}}tr[\partial _{\mu }UU^{\dag
},\partial _{\nu }UU^{\dag }]^{2},  \label{Sk}
\end{equation}%
with the chiral field $U=\exp (2i\boldsymbol{T}\cdot \boldsymbol{\pi }%
/f_{\pi })$ being the nonlinear representation of the Goldstone bosons, $%
U^{\dag }U=1$. Here, $f_{\pi }$ and\ the self-coupling $1/e_{s}$ of the
chiral field scales like $\sqrt{N_{c}}$, while $U$(thereby $U_{5}$) $\sim 
\mathcal{O}(1)$ in the $1/N_{c}$ expansion. We note that the adding\cite%
{Diakonov} of the term $\bar{q}U_{5}q$ in (\ref{chQM}) is used to restore
chiral symmetry $q\rightarrow qe^{2i\gamma ^{5}\boldsymbol{T}\cdot 
\boldsymbol{\theta }}$ under which $U_{5}\rightarrow e^{-2i\gamma ^{5}%
\boldsymbol{T}\cdot \boldsymbol{\theta }}U_{5}e^{-2i\gamma ^{5}\boldsymbol{T}%
\cdot \boldsymbol{\theta }}$.

The model (\ref{chQM}) is invariant under chiral symmetry and can be viewed,
in a sense, as a potential version of the large-$N_{c}$ effective theory\cite%
{Weinberg10} in that instead of using gluon interaction, we use the Hartree
potentials $S$ and $V_{G}$ to effectively describe the flavor-independent
part of the color confining interactions between the valence quarks in
baryon, which are assumed to arise from the gluon interaction at long and
short distance limit, respectively. For the explicit form of $S$ and $V_{G}$
in (\ref{chQM}), we choose them to be of the string-like and
one-gluon-exchange 
\begin{equation}
S=\sigma _{string}r,V_{G}=f_{c}\alpha _{s}(r)/r,  \label{SV}
\end{equation}%
with $\sigma _{string}$ the string tension, $\alpha _{s}(r\gg 1)=g^{2}/4\pi
\sim 1/N_{c}$. Here, $f_{c}=\frac{1}{2}(N_{c}-1/N_{c})$ is the color factor
for the quark-antiquark interaction in its color singlet configuration. It
comes from the formula 
\begin{equation}
f_{c}=\frac{1}{4}tr\left( \frac{\lambda ^{\alpha }}{\sqrt{N_{c}}}\frac{%
\lambda ^{\alpha }}{\sqrt{N_{c}}}\right) =\frac{N_{c}^{2}-1}{2N_{c}},
\label{fc}
\end{equation}%
which$\ $becomes $4/3$ when $N_{c}=3$. Obviously, $f_{c}\alpha _{s}\sim 1$ $+%
\mathcal{O}(1/N_{c}^{2})$ for large $N_{c}$.

For the string tension $\sigma _{string}$ in (\ref{SV}), one known from the $%
SU(N_{c})$ gauge theories on lattice\cite{Lucini} that $\sqrt{\sigma
_{string}}/g_{t}=\mathcal{O}(1)$, up to a $1/N_{c}^{2}$ correction. Thus,
the Hartree potentials (\ref{SV}) scale as $\mathcal{O}(1)$, being
consistent with the argument in \cite{WittenB}. Note that $m_{q}\sim $ $%
M^{B}/N_{c}\sim 1$ since the baryon mass grows linearly with $N_{c}$\cite%
{WittenB}.

We describe our treatment of the strong coupling $\alpha _{s}$ in (\ref{SV})
as follows. With $n_{f}$ quark flavors and the quark masses much less than
the momentum transfer $Q^{2}$, the strong coupling $\alpha _{s}$ is running
with $Q$, in lowest-order QCD, like 
\begin{equation*}
\alpha _{s}(Q^{2})=\frac{12\pi }{(11N_{c}-2n_{f})\ln (Q^{2}/\Lambda
_{QCD}^{2})},
\end{equation*}%
Since $\Lambda _{QCD}\simeq 250MeV$, $\alpha _{s}(Q^{2})$ is small for $%
Q^{2} $ $>1GeV^{2}$, but for any $N_{c}\geq 2$ the above perturbative
formula diverges as $Q\rightarrow \Lambda _{QCD}$, which signals the onset
of confinement. Since in the soft regime we are working we cannot avoid this
divergence, we need to regularize $\alpha _{s}(Q^{2})$, by assuming\cite{CI}%
, for instance, that it saturates at some critical value $\alpha
_{critical}(Q^{2})$ as $Q\rightarrow 0$. In the work \cite{CI}, the running
behavior of $\alpha _{s}(Q^{2})$ has the form of $\sum_{k=1}^{3}\alpha _{k}%
\func{erf}(\gamma _{k}r)$ in the coordinate space through the Fourier
transformation, with $\func{erf}(x)$ the error function. We parametrize this
behavior in a simple form 
\begin{align}
\alpha _{s}(r)& =\alpha _{critical}\tanh \left( \left( r/d\right)
^{2}\right) ,  \label{running} \\
\alpha _{critical}& =\frac{g^{2}}{4\pi }=\frac{g_{t}^{2}}{4\pi N_{c}},
\label{crit}
\end{align}%
with $d$ the infrared cutoff ($\sim 1/\Lambda _{QCD}\simeq 4GeV^{-1}$). The
potential $V_{G}$ in (\ref{SV}), with $\alpha _{s}(r)$ specified by (\ref%
{running}), is shown in \textrm{FIG.1}.

\FRAME{dtbpFU}{3.0813in}{2.1326in}{0pt}{\Qcb{{\protect\small FIG.1. The
chiral angle profile given by (\protect\ref{YY}) and the effective potential 
}$V_{G}(r)${\protect\small \ as a function of radial distance r in fm. }}}{}{%
g12chqm-vy.eps}{\special{language "Scientific Word";type
"GRAPHIC";maintain-aspect-ratio TRUE;display "USEDEF";valid_file "F";width
3.0813in;height 2.1326in;depth 0pt;original-width 6.5078in;original-height
4.4935in;cropleft "0";croptop "1";cropright "1";cropbottom "0";filename
'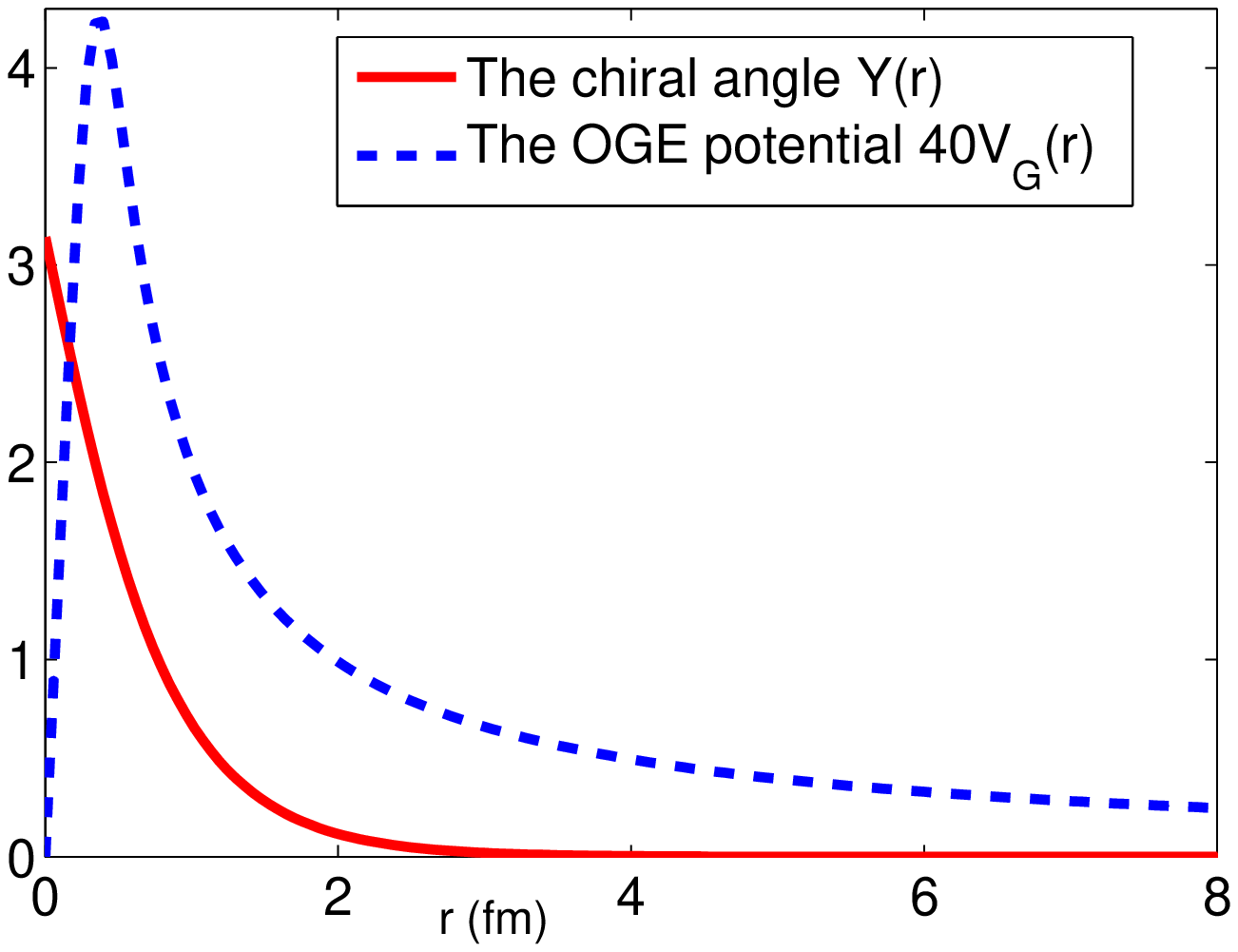';file-properties "XNPEU";}}

Inclusion of the second (Skyrme) term in (\ref{Sk}) may raise issues of
double counting of baryons\footnote{%
This has already been discussed in Ref.\cite{ManoharGeorgi}, where a way is
shown that binding a second pion can be avoided. In Ref. \cite{Weinberg10},
Weinberg argued that the contribution from quark loops in addition to the
tree approximation for purely pionic interactions has to be taken into
account at moderate energies of order $\Lambda _{QCD}$ probed in the
structure of Skyrmions, and this contribution may invalidate the argument of
baryon number double counting.}, since the purely pionic part of the
Lagrangian \ref{chQM} can have Skyrmion solutions, with masses of order $%
N_{c}$\cite{WittenB}, in addition to the $N_{c}$-quark states described
above. Here, the nontrivial solution of the Skyrme Lagrangian (\ref{Sk})
carries a topological charge\cite{Skyrme,Witten83} $B_{\pi }$ known as
baryon number, which may violate the baryon number $B=1$. This is the case
if one ignores the nontrivial vacuum effect that may arise from the spectral
asymmetry of the quark states. However, as Rho et. al. \cite{Rho,GoldstoneW}
noted, the quark spectrum is not symmetric about zero energy and therefore
that the quark vacuum can carry nonzero baryon number, 
\begin{equation}
B_{vac}=-\frac{1}{2}\lim_{t\rightarrow +0}\sum_{n}\mathsf{sgn}%
(E_{n})e^{-t|E_{n}|},  \label{Bvac}
\end{equation}%
where the sum is taken over all positive- and negative energy
single-particle eigenstates of quark with the level $E_{n}$.

While no general proof is available, we argue, by analogy with the proof\cite%
{Goldstone} in chiral bag model, that when applying (\ref{chQM}) to a baryon
the contribution $B_{vac}$ to baryon number from the quark vacuum
polarization (spectrum asymmetry) due to the nontrivial boundary conditions
and the contribution $B_{Sk}$ from the Skyrmion conceals exactly, $%
B_{vac}=-B_{Sk}$,$\ $leaving a unity baryon number in total, $%
B=B_{vac}+B_{Sk}+B_{3q}=B_{3q}=1$. Here, the baryon numbers carried by pion
are defined\cite{Skyrme,GoldstoneW,Witten83} as%
\begin{equation}
B_{\pi }=\int_{V_{\pi }}\frac{d^{3}x}{24\pi ^{2}}\epsilon ^{\mu \nu \lambda
}tr[U^{\dag }\partial _{\mu }UU^{\dag }\partial _{\nu }UU^{\dag }\partial
_{\lambda }U].  \label{Bpion}
\end{equation}%
where $V_{\pi }$ is the region pion field exists. The grounds for the
analogy with the chiral bag comes from the observation that the numbers in (%
\ref{Bvac}) and (\ref{Bpion}) are topological in their nature in the sense
that they are invariant upon smoothly varying of the solutions to the model (%
\ref{chQM}) insofar as the boundary condition for $q$ and $U$ remain intact.
The\ cancellation identity $B_{vac}=-B_{\pi }$, which is valid for the
chiral bag\cite{Goldstone}, remains intact as one smoothly varies the
solution to (\ref{chQM}) to that of the chiral bag, corresponding to varying
of the confining potentials $S$ and $V_{G}$ into that of bag.

\section{Hartree wavefunction of quark in ground state baryon}

The Hartree picture for a baryon suggests a Hamiltonian being the sum of $%
N_{c}$ identical quark Hamiltonians\cite{WittenB},$\mathcal{H}^{Baryon}$ $%
=N_{c}\mathcal{H}^{q}$, and the baryon wavefunction $\Psi (x_{1},\cdots
,x_{N})$ being the product of the single-quark wavefunctions $q(x_{i})$(the
antisymmetric color part is not included here), 
\begin{equation}
\Psi (x_{1},\cdots ,x_{N_{c}})=\prod\nolimits_{i=1}^{N_{c}}q(x_{i}),
\label{baryon}
\end{equation}%
which becomes exact at $N_{c}\rightarrow \infty $ limit\cite{WittenB}. Here, 
$\mathcal{H}^{q}$ is the effective Hamiltonian of a single quark in Hartree
potential, and is taken, in this work, to be the corresponding Hamiltonian
associated with (\ref{chQM}).

To determine $\Psi $, one has to make stationary the variational functional $%
\langle \Psi |\mathcal{H}^{Baryon}-E|\Psi \rangle $, or equivalently, $%
\langle \Psi |\mathcal{H}^{Baryon}-N_{c}\epsilon |\Psi \rangle $ where $%
\epsilon $ is the energy of a single quark in the Hartree potential. If we
denote the Hamiltonian associated with (\ref{chQM}) by $\mathcal{H}^{0}+%
\mathcal{H}^{\chi }$, with $\mathcal{H}^{0}$ the Lagrangian of a single
quark in baryon, the Hartree picture, to the leading order of $N_{c}$,
suggests a Hamiltonian of baryon 
\begin{equation}
\mathcal{H}^{Baryon}=\sum\nolimits_{i=1}^{N_{c}}\mathcal{H}^{0}(x_{i})+%
\mathcal{H}^{\chi },  \label{HchQM}
\end{equation}%
\begin{equation*}
\begin{array}{c}
\mathcal{H}^{(0)}(x_{i})=-i\mathbf{\alpha }\cdot \mathbf{\nabla }_{i}+\gamma
^{0}(m_{q}+S(x_{i}))U_{5}(x_{i})+V_{G}(x_{i}), \\ 
\mathcal{H}^{\chi }=\frac{f{}_{\pi }^{2}}{4}tr(\partial ^{k}U\partial
^{k}U^{\dag })+\frac{1}{16e_{s}^{2}}tr[(\partial ^{k}U^{\dag }\partial
^{k}U)^{2}-(\partial ^{k}U^{\dag }\partial ^{l}U)^{2}],%
\end{array}%
\end{equation*}%
with $S(x)$\textbf{\ }and\textbf{\ }$V_{G}(x)$ given by (\ref{SV}). Here, $%
\mathcal{H}^{\chi }$ is the Hamiltonian associated with Lagrangian(\ref{Sk}%
), $x_{i}$ is the coordinate of the $i$-th quark, while the collective
chiral field $U=U(x)$ depends only on the coordinate $x$ of the Goldstone
boson. Keeping the $N_{c}$-dependence of $f{}_{\pi }$ and $e_{s}$ in mind,
one readily sees that $\mathcal{H}^{\chi }\sim \mathcal{O}(N_{c})$, so that $%
\mathcal{H}^{Baryon}$ $\sim N_{c}$, as argued by Witten\cite{WittenB}.

\ We consider the two-flavor case of baryons with nontrivial configuration
of chiral field of pions. For this, we use the hedgehog ansatz for the
chiral field in (\ref{HchQM}), $U(r)=\exp [iY(r)\mathbf{\tau }\cdot \hat{r}]$%
, where $r$ originated at the mass center of baryons. Writing the quark
wavefunction (with the antisymmetric part of color ignored) as $q=\frac{%
\sqrt{D}}{r}(G(r),-iF(r)\mathbf{\sigma \cdot }\hat{r})^{t}\emph{y}%
_{ljm}(\theta \varphi )\chi _{f}$, with $D\equiv 1/[\int dr(G^{2}+F^{2})]$, $%
\emph{y}_{ljm}$ the\ Pauli spinor and $\chi _{f}$ the flavor wavefunction,
and using (\ref{HchQM}) and (\ref{baryon}), one can rewrite the functional $%
\langle \Psi |\mathcal{H}^{Baryon}-N_{c}\epsilon |\Psi \rangle $ in terms of
the single quark wavefunction $q$. The result is%
\begin{equation}
H^{Baryon}=N_{c}[M^{0}+M^{Sk}/N_{c}],  \label{HB}
\end{equation}%
where 
\begin{equation}
\begin{array}{c}
M^{0}\equiv D^{2}\int dr\left\{ F\frac{dG}{dr}-G\frac{dF}{dr}+\frac{2\kappa 
}{r}GF+(m_{q}+S)\cos Y(G^{2}-F^{2})\right. \\ 
\left. +V_{G}(G^{2}+F^{2})\right\} , \\ 
M^{Sk}=\frac{2\pi f{}_{\pi }}{e_{s}}\int dz\left[ z^{2}\left( \frac{dY}{dz}%
\right) ^{2}+2\sin ^{2}Y\cdot \left( 1+\left( \frac{dY}{dz}\right)
^{2}\right) +\frac{\sin ^{4}Y}{z^{2}}\right]%
\end{array}
\label{H0sk}
\end{equation}%
with $z\equiv r/L\equiv ef_{\pi }r$ and $-\kappa $ the eigenvalue of the
grand spin operator $K=\gamma ^{0}[\mathbf{\Sigma }\cdot (\mathbf{r}\times 
\mathbf{p})+1]$ corresponding to the eigenstate $\emph{y}_{ljm}$.

The equation of motion of (\ref{HB}) for the S-state($l=0$) is 
\begin{equation}
\begin{array}{c}
\frac{dG}{dz}+\frac{\kappa }{z}G=[\varepsilon _{q}+(Lm_{q}+LS)\cos
Y-LV_{G}]F, \\ 
-\frac{dF}{dz}+\frac{\kappa }{z}F=[\varepsilon _{q}-(Lm_{q}+LS)\cos
Y-LV_{G}]G,%
\end{array}
\label{EQM}
\end{equation}%
with $\varepsilon _{q}=L\epsilon $, $LS=z/(e_{s}f_{\pi }a)^{2}$, $%
LV_{G}=f_{c}\alpha _{s}/z$, and $a$ $\equiv 1/\sqrt{\sigma _{string}}$.

Before solving (\ref{EQM}) one has to know $Y(z)$, which is coupled to ($G,F$%
) through the term $\bar{q}U_{5}q\sim (G^{2}-F^{2})\cos Y$ in (\ref{H0sk}).
We use adiabatic approximation to determine the static Skyrmion profile
first and then solve (\ref{EQM}) for ($G,F$). A good fit of the hedgehog
Skyrmion\cite{ANW} was given by that of the kink-like\cite{Atiyah,JiaWang} 
\begin{equation}
Y(z)\simeq 4\arctan (e^{-1.002z}),  \label{YY}
\end{equation}%
which is obtained by minimizing the energy functional $M^{Sk}[Y(r)]$ in (\ref%
{H0sk}) to the value $M^{Sk}=73.62f_{\pi }/e_{s}$, and does not depend upon
the parameters $f_{\pi }$ and $e_{s}\,$. This is comparable to the result $%
M^{Sk}=73f_{\pi }/e_{s}$ in Ref. \cite{ANW}. The plot of the profile (\ref%
{YY}) against $r=z/(e_{s}f_{\pi })$ is presented in \textrm{FIG.1}.

Given the coupling (\ref{running}) and the profile (\ref{YY}), one can solve
the equations (\ref{EQM}) numerically in the case $N_{c}=3$ for the
parameters in \textrm{Table I}. The result for the spinor wavefunction ($G,F$%
) is given in \textrm{FIG.2}.

{\small TABLE I. The parameters of the chiral model. The parameters marked }

{\small by asterisk are inputs from the data, and }${\small \alpha }%
_{critical}{\small =g}_{t}^{2}{\small /(8\pi )}${\small . \ \ \ \ \ \ \ \ \
\ \ \ \ \ \ \ \ \ \ \ \ \ \ \ \ \ \ }

\begin{tabular}[t]{lllll}
\hline\hline
{\small Parameters} & {\small This work} & {\small CI\cite{CI}} & {\small %
ChBM\cite{ChBag}} & {\small SK\cite{ANW}} \\ \hline
${\small N}_{c}$ & ${\small 3}^{\ast }$ &  &  &  \\ 
${\small e}_{s}$ & ${\small 3.80}$ &  & ${\small 4.51}$ & ${\small 5.45}$ \\ 
${\small g}_{t}$ & ${\small 1.57}$ &  &  &  \\ 
${\small m}_{q}{\small =m}_{ud}{\small (MeV)}$ & ${\small 282.8}$ & ${\small %
220}$ & ${\small 330}$ &  \\ 
${\small m}_{s}{\small (MeV)}$ & ${\small 471.3}$ & ${\small 419}$ &  &  \\ 
{\small Zero-point}$(GeV)$ & ${\small -2.1989}$ & ${\small -0.615}$ & $%
0.170( ${\small bag const.}$)$ &  \\ 
{\small Smearing }$\sigma (GeV)$ & ${\small 1.0382}$ & ${\small 1.943}$ &  & 
\\ 
{\small cutoff }$d${\small \ (}$GeV^{-1}${\small )} & ${\small 1.77}$ & $%
{\small 5.00(1/\Lambda }_{QCD}{\small )}$ &  &  \\ 
${\small f}_{\pi }{\small (MeV)}$ & ${\small 93.0}^{\ast }$ &  & ${\small 93}%
^{\ast }$ & ${\small 64.5}$ \\ 
${\small \sigma }_{string}{\small (GeV}^{2}{\small )}$ & ${\small 0.0655}$ & 
${\small 0.15}$ &  &  \\ \hline\hline
\end{tabular}

\FRAME{dtbpFU}{2.7553in}{2.1119in}{0pt}{\Qcb{{\protect\small FIG.2 The
spinor wavefunction (}$G,F${\protect\small ) against the radial distance }$r$%
{\protect\small \ (fm). }}}{}{g12chqmgfsolu.eps}{\special{language
"Scientific Word";type "GRAPHIC";display "USEDEF";valid_file "F";width
2.7553in;height 2.1119in;depth 0pt;original-width 5.834in;original-height
4.3863in;cropleft "0";croptop "1";cropright "1";cropbottom "0";filename
'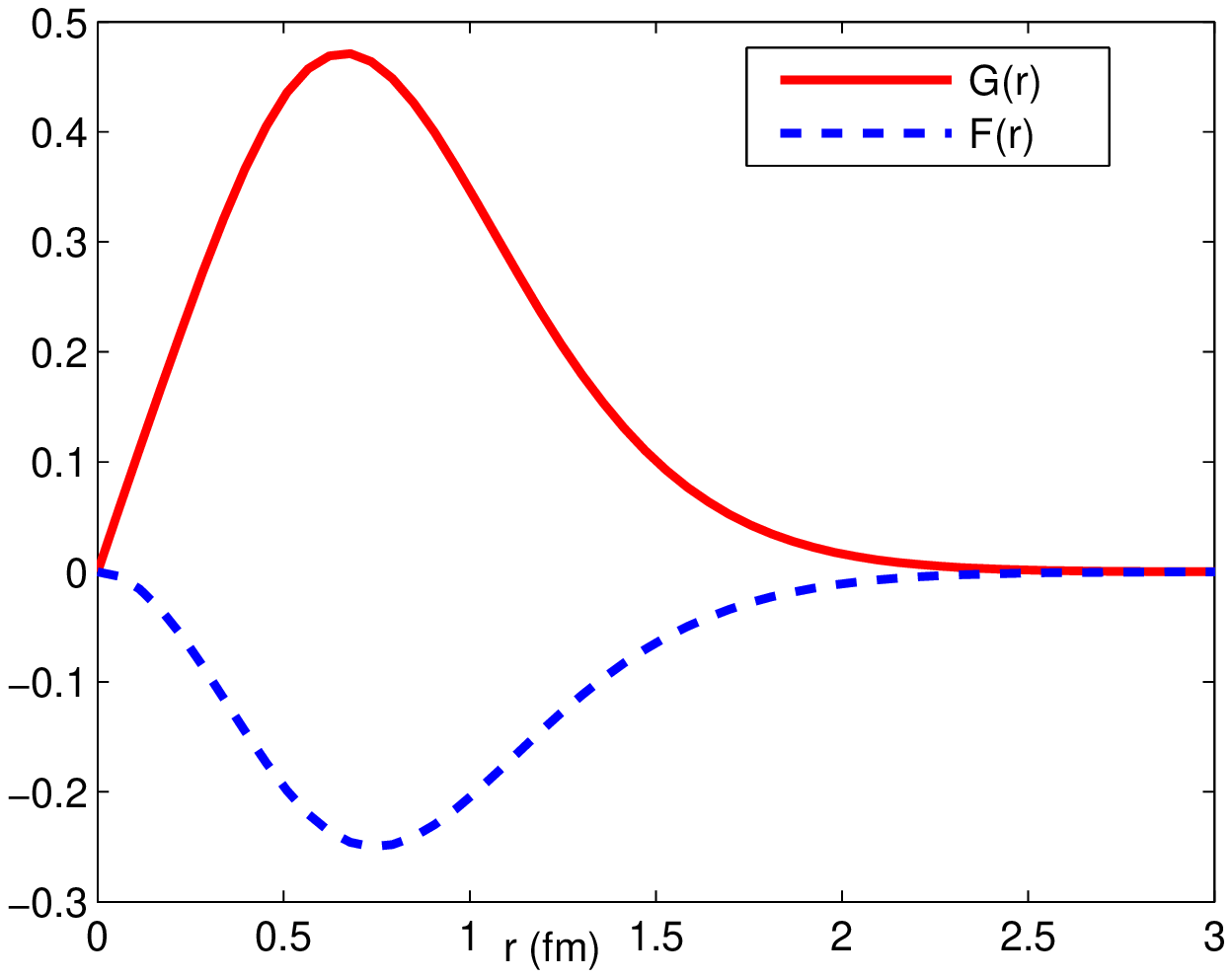';file-properties "XNPEU";}}

\FRAME{dtbpFU}{2.6481in}{1.9657in}{0pt}{\Qcb{{\protect\small FIG.3. The
nonrelativistic wavefunction of the light quark against }$r${\protect\small %
\ (fm).}}}{}{g12chqm-nrfun.eps}{\special{language "Scientific Word";type
"GRAPHIC";display "USEDEF";valid_file "F";width 2.6481in;height
1.9657in;depth 0pt;original-width 6.6366in;original-height 4.3811in;cropleft
"0";croptop "1";cropright "1";cropbottom "0";filename
'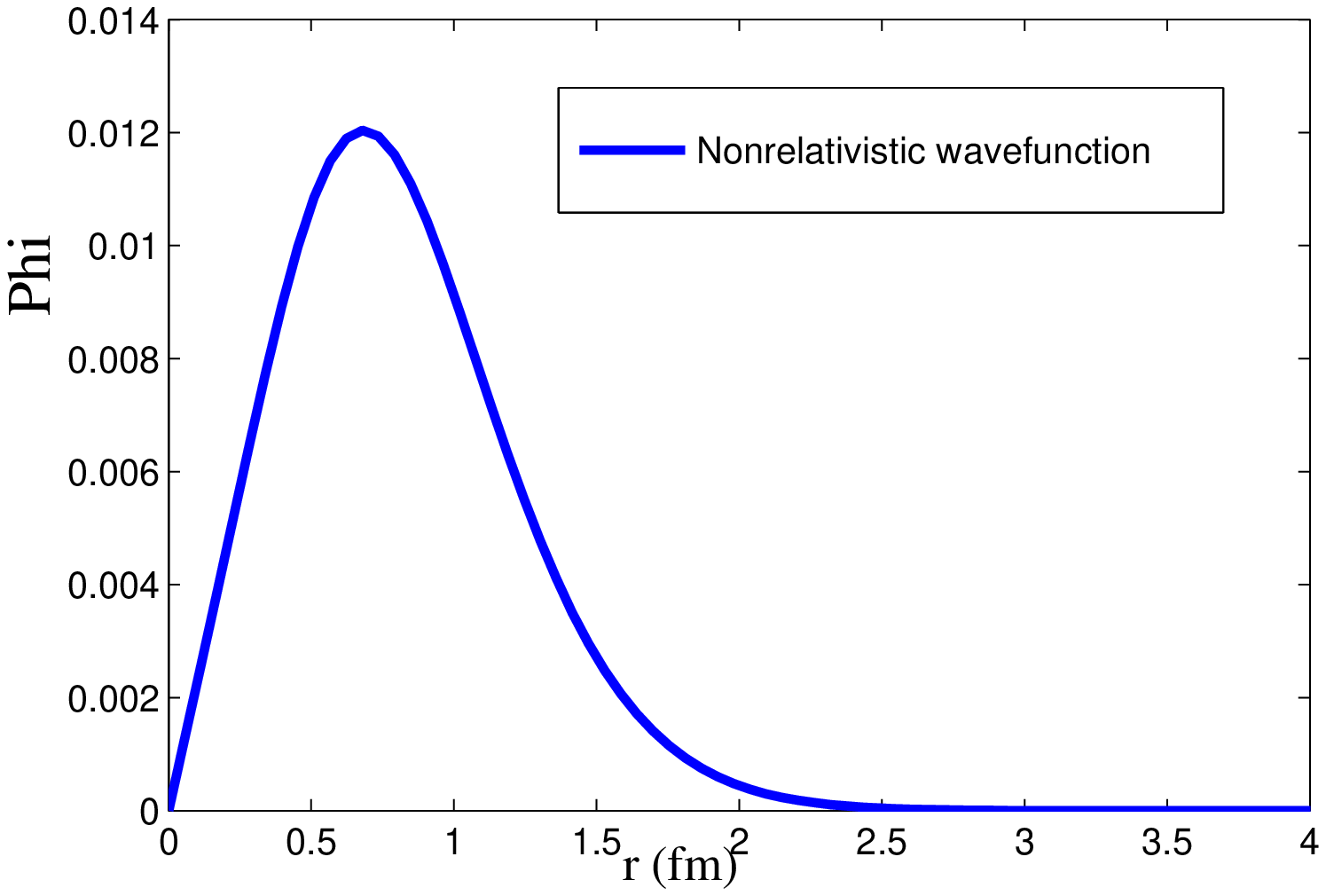';file-properties "XNPEU";}}

It is useful to reduce the spinor wavefunction of quark into its
nonrelativistic form. It can be given by%
\begin{equation}
\phi (\mathbf{x})=\frac{\sqrt{D}G(r)}{r}[1+F(r)^{2}/G(r)^{2}]^{1/2},
\label{PsiNR}
\end{equation}%
which satisfies the normalization $\int |\phi (r)\emph{y}_{ljm}|^{2}d^{3}x=1$%
, and is plotted in \textrm{FIG.3. }

We note that the isospin symmetry is implicitly assumed here so that the
masses of the up and down quarks equal, $m_{q}=m_{ud}$. The parameters are
compared to the corresponding quantities or the counterparts that used in
the other calculations. The solution (\ref{PsiNR}) gives a leading order
approximation to relativistic quark wavefunction. It will be used to compute
the mass correction arising from the strong hyperfine interaction($\sim
1/N_{c}$) in the perturbative framework in the section 4 and 5.

\section{Mass splitting from hyperfine interaction at order $1/N_{c}$}

Being of the order $N_{c}$\cite{WittenB}, the Hartree interactions in (\ref%
{HchQM}) and (\ref{HB}) discussed in the section 2 and 3 is not sufficient
to map the baryon mass splitting in the real world($1/N_{c}=1/3$) which has
to be considered as a subleading correction to the Hartree approximation of
baryon. In fact, with the help of the large $N_{c}$ consistency conditions%
\cite{DashenMano,Jenkin}, Jenkins \cite{Jenkins} shows that baryon mass
splittings are proportional to $J^{2}$($J$ is spin quantum number of baryon)
and first allowed at order $1/N_{c}$. One of striking illustrations for this
is the Skyrme model, where the mass splittings, due to rigid rotation of
soliton, are proportional to $J^{2}/N_{c}$\cite{ANW}.

Following the non-relativistic quark models\cite{DeRujula,CI}, we consider
the spin-dependent pairwise interaction between quarks in baryon which is
inspired by the Breit-Fermi interaction of QED\cite{Sakurai}, and thereby
study the hyperfine mass splitting of baryons. The hyperfine splitting part
of the Breit-Fermi interaction inspired by QED reads\cite{DeRujula} 
\begin{equation}
H_{ij}^{hyp}=\sum_{i<j}\frac{2\alpha _{s}}{3m_{i}m_{j}}\left\{ \frac{8\pi }{3%
}\delta ^{3}(\mathbf{r}_{ij})\mathbf{s}_{i}\cdot \mathbf{s}_{j}+\frac{1}{%
r_{ij}^{3}}\left( 3(\mathbf{s}_{i}\cdot \mathbf{\hat{r}}_{ij})(\mathbf{s}%
_{j}\cdot \mathbf{\hat{r}}_{ij})-\mathbf{s}_{i}\cdot \mathbf{s}_{j}\right)
\right\} .  \label{Hyp}
\end{equation}

To the Hamilton (\ref{HchQM}), which scales like $N_{c}$, we add the
following two-body hyperfine interactions,

\begin{equation}
\mathcal{H}^{ss}=\frac{1}{N_{c}-1}\sum_{i<j=1}^{N_{c}}V_{ij}^{ss},
\label{Hss}
\end{equation}%
with%
\begin{equation}
V_{ij}^{ss}=\frac{f_{c}\alpha _{s}(r_{ij})}{m_{i}m_{j}}\left\{ \frac{8\pi }{3%
}\delta ^{3}(\mathbf{r}_{ij})\mathbf{s}_{i}\cdot \mathbf{s}_{j}+\frac{1}{%
r_{ij}^{3}}\left( 3(\mathbf{s}_{i}\cdot \mathbf{\hat{r}}_{ij})(\mathbf{s}%
_{j}\cdot \mathbf{\hat{r}}_{ij})-\mathbf{s}_{i}\cdot \mathbf{s}_{j}\right)
\right\}  \label{Vss}
\end{equation}%
where $m_{i}$, $\mathbf{s}_{i}$ are the mass and spin of the $i$-th quark in
the baryon center-of-momentum frame, and $\mathbf{r}_{ij}=$ $\mathbf{r}_{i}-%
\mathbf{r}_{j}$ are the relative coordinate of the quark $i$ and $j$. From (%
\ref{Hss}) and (\ref{Vss}), one can check that $\mathcal{H}^{ss}\sim 1/N_{c}$
for the low-spin states with the baryon spin $J\ll N_{c}/2$, if we notice
that the sum in (\ref{Hss}) (put $m_{u}\simeq m_{d}$), 
\begin{equation}
\sum_{i<j}\frac{\mathbf{s}_{i}\cdot \mathbf{s}_{j}}{m_{i}m_{j}}\simeq \frac{1%
}{2m_{u}^{2}}\left[ J(J+1)-\frac{3N_{c}}{4}\right]  \label{ss}
\end{equation}%
becomes $\left[ j(j+1)-3N_{c}\right] /(8m_{u}^{2})$ for $J=j/2\ll N_{c}/2$
and $\left[ N_{c}(N_{c}-1)\right] /(8m_{u}^{2})$ for $J=N_{c}/2$, which
grows like $N_{c}^{2}$ at large $N_{c}$. However, we emphasize that the
later case does not indicate that the perturbative method fully fails for $%
J\sim N_{c}/2$ with $N_{c}=3$ because in that situation (\ref{ss}) is $%
9/(2m_{u})^{2}\sim 1$.

The factor $1/(N_{c}-1)$ in (\ref{Hss}) comes from the generalization of the
"\textit{1/2}" rule\cite{Greenberg,Richard} for pairwise interaction of
quarks in a baryon to the $N_{c}$-body cases. The "\textit{1/2}" rule claims
that the pair potential between two quarks in a baryon is half the
quark-antiquark potential in a meson. Intuitively, it says that two quarks
in a baryon can be seen, when they come closer together, as a localized $%
\bar{3}$ source which is equivalent to an antiquark as in a meson. Sharing
the $\bar{3}$ source by the color charge of each quark, the "1/2" rule
follows. For the same reason, $N_{c}-1$ quarks in a baryon consists of $%
N_{c} $ quarks appear as a $\bar{N}_{c}$ source which is equivalent to an
antiquark, and the factor $1/(N_{c}-1)$ in (\ref{Hss}) follows when the $%
\bar{N}_{c}$ source is shared by each of the $N_{c}-1$ quarks in the $\bar{N}%
_{c}$ source.

The delta function in (\ref{Vss}) is to be smeared\cite{CI} by 
\begin{equation}
\rho _{ij}(\mathbf{r}-\mathbf{r}^{\prime })=\frac{\sigma _{ij}^{3}}{\pi
^{3/2}}e^{-\sigma _{ij}^{2}(\mathbf{r}-\mathbf{r}^{\prime })^{2}}.
\label{smear}
\end{equation}%
While $\sigma _{ij}$ are sensitive to the flavors of quarks for meson, it is
not for baryon\cite{CI}. Thus, for baryon, one can replace $\sigma _{ij}$ by
a single parameter $\sigma $: $\rho _{ij}\rightarrow \rho _{\sigma }\equiv
\sigma ^{3}/\pi ^{3/2}e^{-\sigma ^{2}(\mathbf{r}-\mathbf{r}^{\prime })^{2}}$.

By adding (\ref{Hss}) to (\ref{HB}), one has 
\begin{equation}
M^{Baryon}=N_{c}\left[ M^{0}+\frac{M^{Sk}}{N_{c}}\right] +M^{ss},  \label{MB}
\end{equation}%
\begin{equation}
M^{ss}=\langle B|\mathcal{H}^{ss}|B\rangle .  \label{Mss}
\end{equation}%
The expectation value is taken over the nonrelativistic state $|B\rangle $
of baryon because the operator (\ref{Hss}), analogous to the Breit-Fermi
term of QED, is defined over the subspace which consists of the
nonrelativistic wavefunction(denoted as $\Psi _{B}$). As a nonrelativistic
approximation, $\Psi _{B}$ can be given by%
\begin{equation}
\Psi _{B}\simeq \prod\nolimits_{i=1}^{N_{c}}\phi (x_{i})\chi _{sf},
\label{psiB}
\end{equation}%
with $\chi _{sf}$ the $SU(6)$ spin-flavor wavefunction of baryon. Given $%
\Psi _{B}$ with the one-body orbital wavefunction $\phi $ solved form (\ref%
{EQM}), one has%
\begin{equation}
M^{ss}=\langle \Psi _{B}|\mathcal{H}^{ss}|\Psi _{B}\rangle .  \label{MSS}
\end{equation}

For the S-wave of baryons, the second term in (\ref{Vss}) vanishes. Before
computing (\ref{MSS}) we smear the delta function in (\ref{Vss}),
considering simultaneously the $r$-dependence of $\alpha _{s}(r)$ in (\ref%
{running}), according to the Breit-Fermi-like interaction\cite{CI}, $\frac{7%
}{3m_{i}m_{j}}\mathbf{s}_{i}\cdot \mathbf{s}_{j}\mathbf{\nabla }^{2}V_{int}$%
. This can be done by simply rewriting $\alpha _{s}\delta (\mathbf{r}_{ij})$
in the form of the Laplace expression $\mathbf{\nabla }^{2}\left( -\alpha
_{s}(r_{ij})/4\pi r_{ij}\right) $ at first and then smear $\delta (\mathbf{r}%
_{ij})$ with $\rho _{\sigma }$. Using (\ref{Vss}) one has for the mass
splitting (\ref{MSS})%
\begin{eqnarray}
M^{ss} &=&\sum_{i<j}\frac{8\pi f_{c}\alpha _{critical}}{3(N_{c}-1)m_{i}m_{j}}%
\langle \Psi _{B}|\alpha (r_{ij})\delta ^{3}(\mathbf{r}_{ij})\mathbf{s}%
_{i}\cdot \mathbf{s}_{j}|\Psi _{B}\rangle ,  \notag \\
&=&\sum_{i<j}A_{ij}\frac{\mathbf{s}_{i}\cdot \mathbf{s}_{j}}{m_{i}m_{j}},
\label{Ass}
\end{eqnarray}%
where $\alpha _{s}(r)\equiv \alpha _{critical}\cdot \alpha (r)$ is given by (%
\ref{running}), with%
\begin{equation}
\alpha (r)\equiv \tanh ((r/d)^{2}),  \label{al}
\end{equation}%
and 
\begin{equation}
A_{ij}\equiv \frac{g_{t}^{2}(1+1/N_{c})}{3N_{c}}\left\langle ij\left\vert 
\mathbf{\nabla }^{2}\left( \frac{\alpha (r_{ij})}{4\pi r_{ij}}\right)
\right\vert ij\right\rangle .  \label{AIJ}
\end{equation}%
Here, $|ij\rangle \equiv |\phi (x_{i})\emph{y}_{ljm}^{(i)}\rangle |\phi
(x_{j})\emph{y}_{ljm}^{(j)}\rangle $ with $|\phi (x_{i})\emph{y}%
_{ljm}^{(i)}\rangle $ the spatial and spin wavefunction of the $i$-th quark,
and the relations (\ref{psiB}), (\ref{crit}) and $f_{c}\alpha
_{critical}=g_{t}^{2}/(8\pi )(1-1/N_{c}^{2})$ are used. Noticing the
mathematical identity 
\begin{equation}
\mathbf{\nabla }^{2}\left( \frac{\alpha (r)}{4\pi r}\right) =\alpha
(r)\delta ^{3}(r)-\frac{d^{2}\alpha (r)/dr^{2}}{4\pi r},  \label{Lap}
\end{equation}%
$\,$ and making replacement $\delta ^{3}(r)\rightarrow \rho _{\sigma }(r)$
subsequently, one has the smeared results for (\ref{AIJ}), 
\begin{equation}
A_{ij}=\frac{g_{t}^{2}(1+1/N_{c})}{6N_{c}}[A_{ij}^{(1)}-A_{ij}^{(2)}]\text{.}
\label{Aij}
\end{equation}%
where (see appendix A).%
\begin{eqnarray}
A_{ij}^{(1)} &=&\int \int dr_{1}r_{1}^{2}dr_{2}r_{2}^{2}|\phi
_{i}(r_{1})|^{2}|\phi _{j}(r_{2})|^{2}\frac{\sigma ^{3}e^{-\sigma
^{2}(r_{1}^{2}+r_{2}^{2})}}{\pi ^{3/2}}U_{1}(r_{1},r_{2}),  \label{Aij1} \\
A_{ij}^{(2)} &=&\int \int dr_{1}r_{1}^{2}dr_{2}r_{2}^{2}|\phi
_{i}(r_{1})|^{2}|\phi _{j}(r_{2})|^{2}\frac{1}{4\pi }U_{2}(r_{1},r_{2}),
\label{Aij2}
\end{eqnarray}%
and 
\begin{equation}
\begin{array}{c}
U_{1}(r_{1},r_{2})\equiv \int_{-1}^{+1}due^{2r_{1}r_{2}\sigma ^{2}u}\alpha
\left( (r_{1}^{2}+r_{2}^{2}-2r_{1}r_{2}u)/d^{2}\right) , \\ 
U_{2}(r_{1},r_{2})\equiv \int_{-1}^{+1}du\frac{\alpha ^{^{\prime \prime
}}\left( (r_{1}^{2}+r_{2}^{2}-2r_{1}r_{2}u)/d^{2}\right) }{\sqrt{%
r_{1}^{2}+r_{2}^{2}-2r_{1}r_{2}u}},%
\end{array}
\label{U12}
\end{equation}%
Here, $\alpha (r)$ is given by (\ref{al}) and $\alpha ^{^{\prime \prime
}}=d^{2}\alpha (r)/dr^{2}$ its second-order derivative, and $\phi _{i}$
stands for the solution (\ref{PsiNR}), solved with the mass $m_{i}(i=u,d)$.

Ignoring the isospin breaking, one can replace $A_{ij}^{(m)}$ ($m=1,2$) by a
single parameter $A_{ij}^{(m)}\simeq A^{(m)}$, and thereby finds 
\begin{equation}
A=\frac{g_{t}^{2}(1+1/N_{c})}{6N_{c}}[A^{(1)}-A^{(2)}]\text{.}  \label{A}
\end{equation}%
with $A_{ij}^{(1,2)}$ given by (\ref{Aij1}) and (\ref{Aij2}). From the
equations (\ref{al}), (\ref{Aij1}), (\ref{Aij2}) and (\ref{U12}), one can
easily see that $A\sim 1/N_{c}$ at large $N_{c}$. \ \ \ \ \ \ \ \ \ \ \ \ \
\ \ \ \ \ \ \ \ \ \ \ \ \ \ \ \ \ \ \ \ \ \ \ \ \ \ \ \ \ \ \ 

Putting (\ref{Ass}) into (\ref{HB}), we obtain 
\begin{equation}
M^{Baryon}=N_{c}M^{0}+M^{Sk}+A\sum_{i<j=1}^{N_{c}}\frac{\mathbf{s}_{i}\cdot 
\mathbf{s}_{j}}{m_{i}m_{j}},  \label{mass}
\end{equation}%
with $M^{0},M^{Sk}$ given by (\ref{H0sk}) and $A$ by (\ref{A}). Notice that (%
\ref{mass}) has the form of expansion $M^{Baryon}=a_{0}N_{c}+a_{1}/N_{c}$
since $M^{Sk}\sim N_{c}$ and $A\sim 1/N_{c}$. This is exactly what the large-%
$N_{c}$ QCD predicts\cite{Jenkins} for the $SU(3)_{F}$ symmetry limit.

\section{Extension to three flavor case}

In order to determine the 't Hooft coupling, we need more informations about
the baryons than the two-flavor case considered in preceding sections to fix
the parameters in (\ref{chQM}). For this purpose, one has to extend the
formula (\ref{mass}) into the three flavor case which includes the effects
of the flavor $SU(3)$ breaking. One convenient way to do this is to
attribute the $SU(3)_{F}$ breaking simply, as the naive quark model, to the
mass differences of the quarks between strange and non-strange flavors,
including the resulted differences of the single-particle mean-field energy $%
M^{(i)}$($i=u,d,s$) for quarks, defined by (\ref{H0sk}).

We introduce the strange quark mass $m_{s}$(bigger than $m_{ud}$) so that
the mass difference $m_{s}-m_{ud}$ breaks the flavor $SU(3)$ breaking
explicitly. Instead of computing the strange quark mass $M^{(s)}$ directly
using the bound state approach to strangeness\cite{CaplanKleb}, for
instance, one can determine it by simply fitting the data of the mass
difference between $\Lambda $ and nucleon. We then generalize the mass
formula (\ref{mass}) into 
\begin{equation}
M^{Baryon}=\sum_{i=1}^{N_{c}}M^{(i)}+M^{Sk}+A\sum_{i<j=1}^{N_{c}}\frac{%
\mathbf{s}_{i}\cdot \mathbf{s}_{j}}{m_{i}m_{j}},  \label{Mass}
\end{equation}%
where $A$ and $M^{Sk}$ are taken to be same for three flavors ($i=u,d,s$)
including the strange quark, $M^{(u)}=M^{(u,d)}\equiv M^{0}$\ is defined in (%
\ref{H0sk}). By applying (\ref{Mass}) to $\Lambda =uds$ and nucleon and
eliminating the mass splitting, one has $M^{(s)}=M^{(u)}+\delta M_{\Lambda 
\text{-}N}$, which gives (with the experimental data $\delta M_{\Lambda 
\text{-}N}=175MeV$) 
\begin{equation}
M^{(s)}\simeq M^{(u)}+175MeV.  \label{Ms}
\end{equation}

The relation (\ref{Mass}) is to be compared with the ground-state mass
formula of the nonrelativistic quark model\cite{DeRujula,Gasiorowicz} 
\begin{equation}
M=\sum_{i=1}^{3}m_{i}+A^{\prime }\sum_{i<j=1}^{3}\frac{\mathbf{s}_{i}\cdot 
\mathbf{s}_{j}}{m_{i}m_{j}},  \label{QM}
\end{equation}%
with $A^{\prime }=(2m_{ud})^{2}50MeV$, and that of the Skyrme model\cite%
{Klebanov} 
\begin{equation}
M=M^{soliton}+\left( \frac{1}{2I_{1}}-\frac{1}{2I_{2}}\right) \mathbf{J}^{2}+%
\frac{1}{2I_{2}}\left[ C^{p,q}-\frac{N_{c}^{2}}{12}\right] \text{,}
\label{MSK}
\end{equation}%
with $C^{p,q}$ the Casimir operator and $I_{1,2}\,\sim N_{c}$\ the moments
of inertia of soliton($I_{1}=53/(f_{\pi }e_{s}^{3})$ and $I_{2}=\allowbreak
19.\allowbreak 5/(f_{\pi }e_{s}^{3})$, approximately).

We will show that (\ref{Mass}) and (\ref{QM}) can be understood in the
context of large-$N_{c}$ QCD, which claims \cite{DashenMano,Jenkin} for the $%
SU(3)_{F}$ breaking case,%
\begin{equation}
M=a_{0}N_{c}+a_{1}\frac{\mathbf{J}^{2}}{N_{c}}+b_{1}\epsilon
_{F}T^{8}+\epsilon _{F}b_{2}\frac{J^{i}G^{i8}}{N_{c}}+\mathcal{\cdots }.
\label{MLN}
\end{equation}%
Here, $\epsilon _{F}$ is the $SU(3)_{F}$ breaking, $T^{8}=(N_{c}-N_{s})/%
\sqrt{12}$, $G^{i8}=(J^{i}-3J_{s}^{i})/\sqrt{12}$, $J^{i}$ and $J_{s}^{i}$
are spins of the baryon and strange quark, respectively, and $N_{s}$ the
number of the strange quark, This can be seen if we notice the relation (\ref%
{ss}) for two flavor case and its generalization of the three-flavor case.

The three-flavor generalization of the relation (\ref{ss}) can be given, for
the strangeness$=-N_{s}$ baryons, by 
\begin{equation*}
\begin{array}{c}
\sum_{i<j=1}^{N_{c}}\frac{\mathbf{s}_{i}\cdot \mathbf{s}_{j}}{m_{i}m_{j}}%
=\left( \frac{m_{s}}{m_{q}}-1\right) \sum_{k<k^{\prime }=1}^{N_{c}-N_{s}}%
\frac{\mathbf{s}_{k}\cdot \mathbf{s}_{k^{\prime }}}{m_{q}m_{s}}+\left( 1-%
\frac{m_{s}}{m_{q}}\right) \sum_{I<I^{\prime }=1}^{N_{s}}\frac{\mathbf{s}%
_{I}\cdot \mathbf{s}_{I^{\prime }}}{m_{s}^{2}} \\ 
+\frac{1}{m_{q}m_{s}}\left[ \frac{J(J+1)}{2}-\frac{3N_{c}}{8}\right] ,%
\end{array}%
\end{equation*}%
then, when adding of the extra contribution $N_{s}(M^{(s)}-M^{(u)})$ to the
order-$N_{c}$ mass $N_{c}M^{0}$ stemming from the strange-up energy
difference $\delta M\equiv M^{(s)}-M^{(u)}$, one finds for the baryon mass
formula (\ref{Mass}) with three flavors 
\begin{eqnarray}
M^{Baryon} &=&N_{c}M^{0}+\left[ M^{Sk}-\frac{3(AN_{c})}{8m_{s}m_{q}}\right]
+A\frac{J(J+1)}{2m_{q}m_{s}}+N_{s}\delta M  \notag \\
&&+\left( \frac{m_{s}}{m_{q}}-1\right) A\left[ \sum_{k<k^{\prime
}=1}^{N_{c}-N_{s}}\frac{\mathbf{s}_{k}\cdot \mathbf{s}_{k^{\prime }}}{%
m_{q}m_{s}}-\sum_{I<I^{\prime }=1}^{N_{s}}\frac{\mathbf{s}_{I}\cdot \mathbf{s%
}_{I^{\prime }}}{m_{s}^{2}}\right]  \label{MM}
\end{eqnarray}%
where the small indices $k$ and $k^{\prime }$ refer to the non-strange
quarks($u$ and $d$),\ and the capital indices to the strange quark, $%
m_{q}\equiv m_{ud}$, $\delta M\propto \epsilon _{F}\propto $ $\left( \frac{%
m_{s}}{m_{q}}-1\right) $.

From (\ref{MM}), we see that the third term scales like $1/N_{c}$ the fourth
like $\epsilon _{F}$, and the last scales as $\epsilon _{F}/N_{c}$,
corresponding to the second, the third and the fourth terms in (\ref{MLN}),
respectively. We note here that $\delta M$ is of order $\mathcal{O}(1)$ and $%
A\sim \mathcal{O}(1/N_{c})$. The individual spin product $\mathbf{s}%
_{i}\cdot \mathbf{s}_{j}$ ($i,j=u,d,s$) in (\ref{Mass}) or (\ref{QM}) is $%
3/4 $ if the spins of the quark($i$ and $j$) are parallel and $-3/4$ they
are anti-parallel.

Knowing the wavefunction (\ref{PsiNR}) solved in the section 3 and the
resulted energies $M^{(u)}=M^{0}$ and $M^{Sk}$ given by (\ref{H0sk}), and
using (\ref{Ms}) and (\ref{A}), one can map the masses of the octet and
decuplet baryons using (\ref{Mass}). Our results for the optimal fit of $%
g_{t}$ and $A$ in (\ref{A}) are%
\begin{equation*}
{\small g}_{t}{\small =1.571}\text{{\small , }}{\small A=0.000243,}
\end{equation*}%
with the corresponding masses (\ref{Mass}) calculated for the parameters in 
\textrm{Table I} listed in the \textrm{Table II}.

{\small TABLE II. The calculated baryon masses\ are compared with experiment
(in }$MeV/c^{2}${\small ) when }$N_{c}=3${\small .\ The values for 't Hooft
coupling and mass splitting coefficients in our work are }$g_{t}=1.571$%
{\small \ and }$A=0.000243${\small , respectively. }

\begin{tabular}{cccccc}
\hline\hline
{\small Baryon} & {\small This work} & {\small NQM\cite{CI}} & {\small ChBM%
\cite{ChBag}} & {\small Instance\cite{I}} & {\small Exp.\cite{Ex}} \\ \hline
${\small N}$ & ${\small 939.0}$ & ${\small 960}$ & ${\small 938}$ & ${\small %
938}$ & ${\small 938.9}$ \\ 
$\Lambda $ & ${\small 1114.0}$ & ${\small 1115}$ & ${\small 1149}$ & $%
{\small 1116}$ & ${\small 1115.7}$ \\ 
$\Sigma $ & ${\small 1117.4}$ & ${\small 1190}$ & ${\small 1211}$ & ${\small %
1184}$ & ${\small 1193.1}$ \\ 
$\Xi $ & ${\small 1291.0}$ & ${\small 1305}$ & ${\small 1369}$ & ${\small %
1329}$ & ${\small 1318.1}$ \\ 
$\Delta $ & ${\small 1232.0}$ & ${\small 1230}$ & ${\small 1205}$ & ${\small %
1239}$ & ${\small 1232.0}$ \\ 
$\Sigma ^{\ast }$ & ${\small 1405.3}$ & ${\small 1370}$ & ${\small 1370}$ & $%
{\small 1383}$ & ${\small 1384.6}$ \\ 
$\Xi ^{\ast }$ & ${\small 1580.3}$ & ${\small 1505}$ & ${\small 1531}$ & $%
{\small 1528}$ & ${\small 1533.4}$ \\ 
${\small \Omega }$ & ${\small 1752.9}$ & ${\small 1635}$ & ${\small 1683}$ & 
${\small 1672}$ & ${\small 1672.5}$ \\ \hline\hline
\end{tabular}

\section{Summary and concluding remark}

We studied the 't Hooft coupling and the baryon masses hyperfine splittings
of the ground state baryons in a chiral quark model inspired by the large-$%
N_{c}\ $QCD. The Hartree picture of baryon at large $N_{c}$ is used to map
the masses of octet and decuplet baryons in the ground states. The 't Hooft
coupling $g_{t}$ that reproduces data of masses is found to be about $1.57$.
We believe that this gives an indication that in Weinberg's effective field
theory, the `t Hooft coupling $g_{t}$ becomes small in the sense that $%
g_{t}^{2}/4\pi N_{c}\ll 1$ at moderate energies.

The fact that the quantum numbers and the group theoretic structure of
baryons in the Skyrme model is identical to that of the naive quark model at
large $N_{c}$\cite{Manohar} implies some yet unknown connection between the
two. We hope that the large-$N_{c}$ chiral model considered in this paper
will enhance this connection, as indicated by (\ref{Mass}), (\ref{QM}) and (%
\ref{MSK}), in the sense that both fit well into the large $N$ formula (\ref%
{MLN}).

\begin{description}
\item[\textbf{Acknoledgements}] 
\end{description}

D. J thanks X.Q Li, D.Y Chen and X. Liu for useful discussions. This work is
supported by the Natural Science Foundation of China(No.11265014).

\begin{description}
\item[Appendix A] 
\end{description}

We write the smeared matrix element $\left\langle ij\left\vert \alpha
(r_{ij})\rho _{\sigma }(r_{ij})\right\vert ij\right\rangle $ in (\ref{AIJ})
as 
\begin{align*}
& \int \int d^{3}r_{1}d^{3}r_{2}|\phi _{i}(r_{1})|^{2}|\phi _{j}(r_{2})|^{2}|%
\emph{y}_{ljm}^{(i)}\emph{y}_{ljm}^{(j)}|^{2}\alpha (r_{ij})\rho _{\sigma
}(r_{ij}) \\
& =\frac{(2\pi )^{2}}{(4\pi )^{2}}\int_{0}^{\pi }d\theta _{1}\sin (\theta
_{1})d\theta _{2}\sin (\theta _{2})\int dr_{1}dr_{2}r_{1}^{2}r_{2}^{2}|\phi
_{i}(r_{1})|^{2}|\phi _{j}(r_{2})|^{2} \\
& \times \alpha (|\mathbf{r}_{1}-\mathbf{r}_{2}|)\rho _{\sigma }(|\mathbf{r}%
_{1}-\mathbf{r}_{2}|),
\end{align*}%
where $r_{ij}^{2}=|\mathbf{r}_{1}-\mathbf{r}%
_{2}|^{2}=r_{1}^{2}+r_{2}^{2}-2r_{1}r_{2}\cos (\theta _{2}-\theta _{1})$
when one choose the frame ($r_{1},\theta _{1},\varphi _{1}$) and ($%
r_{1},\theta _{1},\varphi _{1}$) so that the three vector $\mathbf{r}_{1},%
\mathbf{r}_{2},\mathbf{z}$ in the same plane. Make a change of variables
from ($r_{1},\theta _{1},\varphi _{1}$) and ($r_{1},\theta _{1},\varphi _{1}$%
) to ($r_{1},\theta _{1},\varphi _{1}$) and ($r_{1},\theta _{2}-\theta
_{1},\varphi _{2}$), and performing the integration, one gets (\ref{Aij1}),
with $U_{1}(r_{1},r_{2})$ given by (\ref{U12}). Similarly, one can get (\ref%
{Aij2}) for $\left\langle ij\left\vert \alpha (r_{ij})\rho _{\sigma
}(r_{ij})\right\vert ij\right\rangle $. Here, we give the alternative
expressions for (\ref{Aij1}), (\ref{Aij2}) and (\ref{U12}), with the
integration over the dimensionless variable $z=r/L$, which are convenient
for the numerical computation of mass splittings.

\begin{align*}
A_{ij}^{(1)}& =\frac{(DL)^{2}\sigma ^{3}}{\pi ^{3/2}}\int \int
dz_{1}dz_{2}|\phi _{i}(z_{1})|^{2}|\phi _{j}(z_{2})|^{2}e^{-L^{2}\sigma
^{2}(z_{1}^{2}+z_{2}^{2})}U_{1}(z_{1},z_{2}) \\
A_{ij}^{(2)}& =\frac{(DL)^{2}}{4\pi }\int \int dz_{1}dz_{2}|\phi
_{i}(z_{1})|^{2}|\phi _{j}(z_{2})|^{2}U_{2}(z_{1},z_{2})
\end{align*}%
\begin{align*}
U_{1}(z_{1},z_{2})& =\int_{-1}^{+1}due^{2z_{1}z_{2}L^{2}\sigma ^{2}u}\tanh
\left( (z_{1}^{2}+z_{2}^{2}-2z_{1}z_{2}u)\left( \frac{L}{d}\right)
^{2}\right) , \\
U_{2}(z_{1},z_{2})& =\frac{1}{L^{3}}\int_{-1}^{+1}du\frac{\tanh ^{^{\prime
\prime }}\left( (z_{1}^{2}+z_{2}^{2}-2z_{1}z_{2}u)\left( \frac{L}{d}\right)
^{2}\right) }{\sqrt{z_{1}^{2}+z_{2}^{2}-2z_{1}z_{2}u}}.
\end{align*}

\end{document}